\documentclass[aps,prl,twocolumn,superscriptaddress,floatfix]{revtex4}%
\usepackage{graphicx}
\usepackage{dcolumn}
\usepackage{bm}
\usepackage{hyperref}
\usepackage{amssymb}
\usepackage{amsmath}
\usepackage{amsfonts}
\usepackage{amssymb}
\usepackage{color}%
\begin{document}
\title{Observation of photoassociation resonances in ultracold atom-molecule collisions}

\author{Jin Cao}
\thanks{These authors contributed equally to this work.}
\affiliation{Hefei National Research Center for Physical Sciences at the Microscale and School of Physical Sciences, University of Science and Technology of China, Hefei 230026, China}
\affiliation{Shanghai Research Center for Quantum Science and CAS Center for Excellence in Quantum Information and Quantum Physics, University of Science and Technology of China, Shanghai 201315, China}
\author{Bo-Yuan Wang}
\thanks{These authors contributed equally to this work.}
\affiliation{Hefei National Research Center for Physical Sciences at the Microscale and School of Physical Sciences, University of Science and Technology of China, Hefei 230026, China}
\affiliation{Shanghai Research Center for Quantum Science and CAS Center for Excellence in Quantum Information and Quantum Physics, University of Science and Technology of China, Shanghai 201315, China}
\author{Huan Yang}
\thanks{These authors contributed equally to this work.}
\affiliation{Hefei National Research Center for Physical Sciences at the Microscale and School of Physical Sciences, University of Science and Technology of China, Hefei 230026, China}
\affiliation{Shanghai Research Center for Quantum Science and CAS Center for Excellence in Quantum Information and Quantum Physics, University of Science and Technology of China, Shanghai 201315, China}
\author{Zhi-Jie Fan}
\affiliation{Hefei National Research Center for Physical Sciences at the Microscale and School of Physical Sciences, University of Science and Technology of China, Hefei 230026, China}
\affiliation{Shanghai Research Center for Quantum Science and CAS Center for Excellence in Quantum Information and Quantum Physics, University of Science and Technology of China, Shanghai 201315, China}

\author{Zhen Su}
\affiliation{Hefei National Research Center for Physical Sciences at the Microscale and School of Physical Sciences, University of Science and Technology of China, Hefei 230026, China}
\affiliation{Shanghai Research Center for Quantum Science and CAS Center for Excellence in Quantum Information and Quantum Physics, University of Science and Technology of China, Shanghai 201315, China}
\author{Jun Rui}
\affiliation{Hefei National Research Center for Physical Sciences at the Microscale and School of Physical Sciences, University of Science and Technology of China, Hefei 230026, China}
\affiliation{Shanghai Research Center for Quantum Science and CAS Center for Excellence in Quantum Information and Quantum Physics, University of Science and Technology of China, Shanghai 201315, China}
\affiliation{Hefei National Laboratory, University of Science and Technology of China, Hefei 230088, China}
\author{Bo Zhao}
\affiliation{Hefei National Research Center for Physical Sciences at the Microscale and School of Physical Sciences, University of Science and Technology of China, Hefei 230026, China}
\affiliation{Shanghai Research Center for Quantum Science and CAS Center for Excellence in Quantum Information and Quantum Physics, University of Science and Technology of China, Shanghai 201315, China}
\affiliation{Hefei National Laboratory, University of Science and Technology of China, Hefei 230088, China}
\author{Jian-Wei Pan}
\affiliation{Hefei National Research Center for Physical Sciences at the Microscale and School of Physical Sciences, University of Science and Technology of China, Hefei 230026, China}
\affiliation{Shanghai Research Center for Quantum Science and CAS Center for Excellence in Quantum Information and Quantum Physics, University of Science and Technology of China, Shanghai 201315, China}
\affiliation{Hefei National Laboratory, University of Science and Technology of China, Hefei 230088, China}

\begin{abstract}{Photoassociation of ultracold atoms is a resonant light-assisted collision process, in which two colliding atoms absorb a photon and form an excited molecule. Since the first observation about three decades ago, the photoassociation of ultracold atoms has made a significant impact on the study of ultracold atoms and molecules. Extending the photoassociation of atoms to the photoassociation of atom-molecule pairs or molecule-molecule pairs will offer many new opportunities in the study of precision polyatomic molecular spectroscopy, formation of ultracold polyatomic molecules, and quantum control of molecular collisions and reactions. However, the high density of states and the photoexcitation of the collision complex by the trapping laser make photoassociation into well-defined quantum states of polyatomic molecules extremely difficult. Here we report on the observation of photoassociation resonances in ultracold collisions between $^{23}$Na$^{40}$K molecules and $^{40}$K atoms. We perform photoassociation in a long-wavelength optical dipole trap to form deeply bound triatomic molecules in the electronically excited states. The atom-molecule Feshbach resonance is used to enhance the free-bound Franck-Condon overlap. The photoassociation into well-defined quantum states of excited triatomic molecules is identified by observing resonantly enhanced loss features. These loss features depend on the polarization of the photoassociation lasers, allowing us to assign the rotational quantum numbers. The observation of ultracold atom-molecule photoassociation resonances paves the way toward preparing ground-state triatomic molecules, provides a new high-resolution spectroscopy technique for polyatomic molecules, and is also important to atom-molecule Feshbach resonances.}
\end{abstract}
\maketitle

Photoassociation of ultracold atoms is a resonant light-assisted collision process, in which two colliding atoms absorb a photon and form an excited molecule if the laser frequency is in resonance with a free-bound transition \cite{Lett1995,Stwalley1999,Jones2006}. Since the first experimental observation  about three decades ago \cite{Thorsheim1987,Lett1993,Miller1993}, the photoassociation of ultracold atoms has made significant contributions to the study of ultracold atoms and molecules. For example, photoassociation is a general approach to forming molecules. In contrast to magnetoassociation which forms extremely weakly bound Halo molecules \cite{Koehler2006}, photoassociation can form molecules in various rovibrational levels, ranging from long-range excited molecules to tightly bound ground-state molecules \cite{Fioretti1998,Gabbanini2000,Wang2004,Mancini2004,Sage2005,Deiglmayr2008,Aikawa2010,Reinaudi2012}, by tuning the laser frequency  and using multiphoton processes. In addition to forming molecules, photoassociation provides a new high-resolution molecular spectroscopy technique \cite{Lett1995,Stwalley1999,Jones2006,Thorsheim1987,Lett1993,Miller1993}, which can study the high-lying states that are difficult to access for conventional molecular spectroscopy. Photoassociation can also be used as an optical Feshbach resonance to control the ultracold collisions between atoms \cite{Fedichev1996,Theis2004}, which is particularly useful to the atoms that do not have magnetic Feshbach resonances. Besides, photoassociation has been used as a powerful tool for studying molecular Bose-Einstein condensates \cite{Wynar2000}, ultralong-range Rydberg macromolecules \cite{Bendkowsky2009,Rui2019}, and ultracold reactions in optical tweezers \cite{LiuLR2018}.

With the preparation of ultracold diatomic molecules \cite{Ni2008,Molony2014,Takekoshi2014,Park2015,Guo2016,Rvachov2017,Voges2020,Stevenson2023,LiuLR2018,He2020,DeMarco2019,Schindewolf2022,Cao2023}, the atom-atom photoassociation may be extended to the atom-molecule photoassociation or the molecule-molecule photoassociation. Such extensions will offer many new research opportunities in the study of precision polyatomic molecular spectroscopy, the formation of ultracold polyatomic molecules, scattering resonances in ultracold molecular collisions, and quantum control of light-assisted ultracold reactions. Extending the photoassociation from a diatomic system to a polyatomic system has been proposed since about two decades ago \cite{Stwalley1999,Jones2006}. However, the complexity of the polyatomic molecules makes such extensions extremely challenging. Recently, the prospects of ultracold photoassociation in atom-molecule collisions \cite{Lepers2010,Olivier2015,Schnabe2021,Shammout2023,Elkamshishy2023} or molecule-molecule collisions \cite{Gacesa2021} have been theoretically studied, discussing the feasibility to observe photoassociation resonances in ultracold collisions involving molecules and to create stable ultracold polyatomic molecules. Because the short-range interactions in atom-molecule collisions or molecule-molecule collisions are very complicated, only the long-range polyatomic states near the dissociation limit can be studied. However, the density of states of these high-lying polyatomic molecular states is very high owing to the heavy mass of atoms and the strongly anisotropic interactions \cite{Mayle2012, Mayle2013}. Consequently, it is unclear whether the photoassociation will result in well-resolved resonant features or a broad quasi-continuous profile containing numerous unresolvable resonances \cite{Olivier2015,Shammout2023}. Another major obstacle to the observation of photoassociation resonances is that the trapping laser can excite the polyatomic collision complex with a high excitation rate \cite{Christianen2019,Gregory2020,Liu2020,Nichols2022}. The photoexcitation by the trapping laser causes a short-range loss with a near-unity probability and thus may hinder the possibility of photoassociation into well-defined quantum states of polyatomic molecules.

\begin{figure}[ptb]
\centering
\includegraphics[width=8cm]{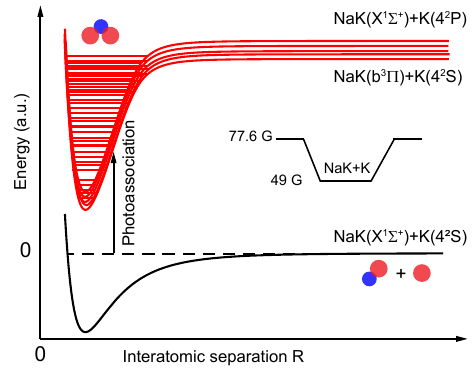}
\caption{Illustration of the photoassociation in ultracold collisions between $^{23}$Na$^{40}$K molecules and $^{40}$K atoms. The ultracold $^{23}$Na$^{40}$K molecule and $^{40}$K atom absorb a photon and form a triatomic molecule in the electronically excited state if the laser frequency is in resonance with a free-bound transition.  To suppress the photoexcitation of the collision complex by the trapping laser and to avoid the high density of states near the dissociation limit, we perform photoassociation to form deeply bound excited molecules in a long-wavelength optical dipole trap. To enhance the free-bound Franck-Condon overlap, photoassociation is performed in the vicinity of the atom-molecule Feshbach resonance by ramping the magnetic field to about 49 G, which is slightly above the resonance located at about 48.2 G.} %
\label{fig1}%
\end{figure}

Here, we report on the observation of photoassociation resonances in ultracold collisions between $^{23}$Na$^{40}$K molecules and $^{40}$K atoms. To suppress the photoexcitation of the collision complex by the trapping laser, we study photoassociation in a long-wavelength optical dipole trap. To overcome the difficulty caused by the high density of states, we use photoassociation to form deeply bound triatomic molecules in electronically excited states. The free-bound Franck-Condon overlap is enhanced \cite{Junker2008,Pellegrini2008} by using an atom-molecule Feshbach resonance \cite{Yang2019,Son2022}. We identify the photoassociation of $^{23}$Na$^{40}$K molecules and $^{40}$K atoms into well-defined quantum states of excited triatomic molecules by observing resonantly enhanced loss features. The loss features depend on the polarization of the photoassociation lasers, allowing us to assign the rotational quantum number of the triatomic molecular states. The observation of photoassociation resonances in ultracold atom-molecule collisions paves the way toward preparing ground-state triatomic molecules, provides a new high-resolution spectroscopy technique for polyatomic molecules, and is also important to atom-molecule Feshbach resonances.

Our experiment starts with the preparation of an ultracold mixture of $^{23}$Na$^{40}$K molecules and $^{40}$K atoms in a long-wavelength optical dipole trap. The experimental procedures are given in Methods. In brief, we first create a deeply degenerate atomic mixture of $^{23}$Na and $^{40}$K atoms with a large number imbalance in a large-volume optical dipole trap at a wavelength of 1064 nm \cite{Cao2023}. We then load the atoms into a long-wavelength crossed-beam optical dipole trap formed by two laser beams.  The wavelengths of the two trapping beams are 1558 nm and 1583 nm, respectively. The horizontal trapping beam propagates along the direction of the magnetic field and the vertical trapping beam propagates perpendicular to the direction of the magnetic field.  After loading the atomic mixture into the long-wavelength optical dipole trap, we create $^{23}$Na$^{40}$K Feshbach molecules at 77.6 G by magnetoassociation and then transfer them into the rovibrational ground state by stimulated Raman adiabatic passage. After removing the $^{23}$Na atoms by a resonant light pulse, we obtain an ultracold mixture of $^{23}$Na$^{40}$K molecules and $^{40}$K atoms. The $^{23}$Na$^{40}$K molecules are prepared in the maximally polarized hyperfine state of the rovibrational ground state and the $^{40}$K atoms are in the lowest hyperfine state. The number of $^{40}$K atoms is about one order of magnitude larger than that of $^{23}$Na$^{40}$K molecules.

\begin{figure}[ptb]
\centering
\includegraphics[width=8cm]{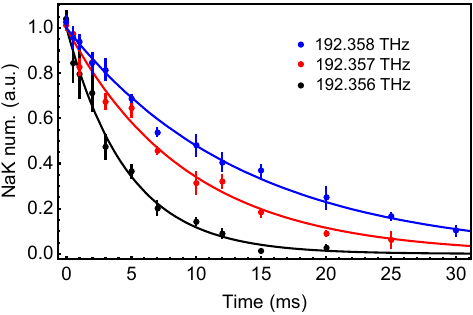}
\caption{The decay of $^{23}$Na$^{40}$K molecules in the atom-molecule mixture for different photoassociation laser frequencies. The time evolution of $^{23}$Na$^{40}$K molecules is recorded for different photoassociation laser frequency. The 1558 nm trapping laser is used as the photoassociation laser, the polarization of which is set to $\sigma^{+}$. The solid lines are exponential fits to the data with $1/e$ lifetimes of 13.6(5) ms at 192.358 THz, 9.3(6) ms at 192.357 THz, and 4.7(3) ms at 192.356 THz.  Each point represents the average of 3-5 measurements and error bars represent the standard error of the mean.}%
\label{fig2}%
\end{figure}

In the long-wavelength optical dipole trap, we use the horizontal trapping laser as the photoassociation laser. The experimental scheme is illustrated in Fig. 1. The 1558 and 1583 nm photoassociation lasers may form deeply bound excited molecules below the NaK($b^{3}\Pi)+$ K(4$^{2}$S) dissociation limit by about 5100 cm$^{-1}$. The density of states of these deeply bound states is much lower than that of high-lying states near the dissociation limit, and thus there may exist well-resolved triatomic molecules that can be addressed by photoassociation. However, the free-bound Franck-Condon overlap for the deeply-bound states is very low. To enhance the Franck-Condon overlap, we study photoassociation in the vicinity of an atom-molecule Feshbach resonance. In our experiment, after preparing the atom-molecule mixture at 77.6 G, we ramp the magnetic field to about 49 G, which is slightly higher than the resonance position. We hold the $^{23}$Na$^{40}$K molecules and $^{40}$K atoms for a certain time and then remove the $^{40}$K atoms using a resonant light pulse. After that, we ramp the magnetic field back to 77.6 G and measure the number of remaining $^{23}$Na$^{40}$K molecules by transferring them into Feshbach states for detection.

During the hold time, the inelastic collisions between $^{23}$Na$^{40}$K molecules and $^{40}$K atoms lead to the loss of $^{23}$Na$^{40}$K molecules.
We change the frequency of the horizontal trapping laser and use the decay of the $^{23}$Na$^{40}$K molecules as a probe of the photoassociation. If the trapping laser frequency is in resonance with a free-bound transition, a pair of $^{23}$Na$^{40}$K molecule and $^{40}$K atom will absorb a photon and form an excited triatomic molecule, which will decay due to spontaneous emission. Consequently, the decay of $^{23}$Na$^{40}$K molecules will be resonantly enhanced. Therefore, the photoassociation into a well-defined quantum state of triatomic molecules will manifest itself as a resonantly enhanced loss feature.

However, it is unclear whether the enhanced loss caused by the photoassociation can be distinguished from the background loss. This is because in the vicinity of the atom-molecule Feshbach resonance, the decay of the $^{23}$Na$^{40}$K molecules is largely enhanced. At about 49 G, the typical lifetime of the $^{23}$Na$^{40}$K molecules in the atom-molecule mixture in the long-wavelength optical dipole trap is about 10 ms. Various mechanisms may cause the losses. One major loss mechanism is the photoexcitation of the collision complex by the trapping laser. Therefore, to observe the resonant loss feature, the enhanced loss caused by the photoassociation resonances must be larger than the loss caused by the photoexcitation of the collision complex by the trapping laser. Previous works have demonstrated that in atom-molecule collisions the photoexcitation by the 1064 nm trapping laser can cause a short-range loss with a near-unity probability \cite{Nichols2022}. Although it is proposed that a long-wavelength optical dipole trap may be helpful to suppress the photoexcitation of the collision complex \cite{Christianen2019}, it is unclear whether such suppression is sufficient for the observation of resonant features. In our experiment, we first carefully study the decay of $^{23}$Na$^{40}$K molecules at different photoassociation laser frequencies. We find that in some frequency regions, the decay rate of the $^{23}$Na$^{40}$K molecules can be enhanced by a factor of about 3 via varying the laser frequency. An example of such an enhanced loss is shown in Fig. 2, where the 1558 nm laser is used as the photoassociation laser. The decay rate of the $^{23}$Na$^{40}$K molecules at 192.356 THz (near a photoassociation resonance) is approximately 3 times larger than the decay rate at 192.358 THz (far away from a photoassociation resonance). We attribute such an enhanced loss to the photoassociation of $^{23}$Na$^{40}$K molecules and $^{40}$K atoms into a well-defined quantum state of triatomic molecules.

\begin{figure}[ptb]
\centering
\includegraphics[width=8cm]{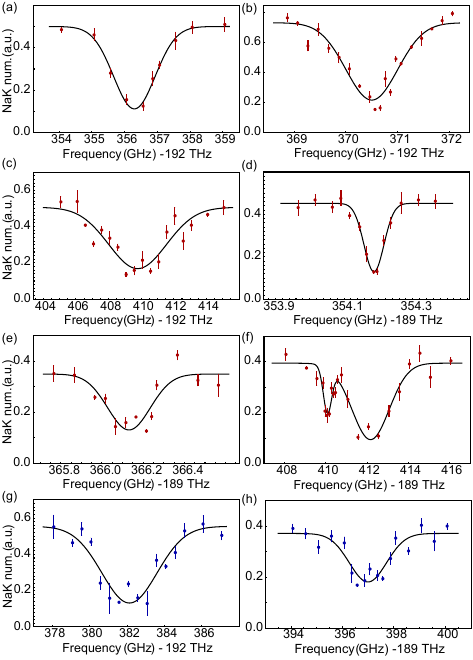}
\caption{Observation of photoassociation resonances in ultracold collisions between $^{23}$Na$^{40}$K molecules and $^{40}$K atoms. The remaining number of $^{23}$Na$^{40}$K molecules is plotted as a function of the frequency of the photoassociation laser. These resonant loss features provide clear evidence for the photoassociation into well-defined quantum states of excited triatomic molecules. (a)-(f), the photoassociation laser is $\sigma^{+}$ polarized. (g) and (h), the photoassociation laser is $\sigma^{-}$ polarized. The solid lines are Gaussian fits to the data points. Each point represents the average of 3-6 measurements and error bars represent the standard error of the mean. }%
\label{fig3}%
\end{figure}

To search for more resonant loss features in a large frequency range, we record the number of remaining $^{23}$Na$^{40}$K molecules for a fixed hold time of 10 ms as a function of the horizontal trapping beam frequency, normalized to that for a zero hold time. The polarization of the horizontal trapping beam is set to be either $\sigma^{+}$ or $\sigma^{-}$. We first use the 1558 nm laser as the photoassociation laser to search in the frequency range between 192.35 and 192.42 THz. We then exchange the two trapping lasers and use the 1583 nm laser as the photoassociation laser to search in the range between 189.33 and 189.42 THz. In total, nine resonantly enhanced loss features are observed. The results are shown in Fig. 3. We find that these loss features are dominantly excited by either $\sigma^{+}$ or $\sigma^{-}$ photoassociation laser.  The dependence on the polarization of the photoassociation laser indicates that these resonant loss features represent rotationally resolved molecules. The data are fit to Gaussian functions. The center frequencies and the widths obtained by Gaussian fits are given in Tab. 1. The loss feature located at 189.35419 THz has a width of about 40 MHz, which is much narrower than other loss features.

\begin{table} [pth]
\centering
\begin{tabular}{c c c c c}
\hline \hline

  & Frequency (THz) & Width (GHz) & Polarization & $J$    \\ \hline

1 &192.3563 & 0.90 & $\sigma^{+}$ & 1/2       \\
2 &192.3705 & 0.71 & $\sigma^{+}$  & 1/2       \\
3 &192.4098 & 2.5  & $\sigma^{+}$  & 1/2      \\
4 &189.35419 & 0.04 & $\sigma^{+}$  & 1/2         \\
5 &189.3661 & 0.15 &$\sigma^{+}$  & 1/2          \\
6 &189.4101 & 0.29 & $\sigma^{+}$  & 1/2         \\
7 &189.4122 & 1.3 & $\sigma^{+}$  & 1/2         \\
8 &192.3821 & 2.1 & $\sigma^{-}$  & 3/2        \\
9 &189.3970 & 1.0 &$\sigma^{-}$  & 3/2         \\

\hline \hline
\end{tabular}
\caption{The central frequencies and widths are obtained by Gaussian fits to the resonant loss features. The polarization of the photoassociation light is either $\sigma^{+}$ or $\sigma^{-}$. The total rotational quantum numbers of these excited molecular states are assigned. }
\label{Table}
\end{table}

These well-resolved resonant loss features represent the photoassociation of $^{23}$Na$^{40}$K molecules and $^{40}$K atoms into well-defined quantum states of $^{23}$Na$^{40}$K$_2$  molecules in the electronically excited state. To gain further understanding of these molecular states, we calculate the energy of the equilibrium states of the doublet electronic excited states of $^{23}$Na$^{40}$K$_{2}$ molecules using \emph{ab initio} calculations (see details in Methods). These equilibrium states correspond to the local minimum of the three-dimensional potential energy surface. The energies of the lowest four equilibrium states above the NaK(X$^{1}\Sigma^{+})+$ K(4$^{2}$S) dissociation limit are shown in Fig. 4. The $3^{2}A_{1}$, $2^{2}B_{1}$, and $1^{2}B_2$ states have $C_{2v}$ symmetry, while the $3^{2}A'$ state has $C_s$ symmetry. The molecular states observed in our experiment are lower than the $3^{2}A_{1}$ and $2^{2}B_{1}$ states, while they are higher than the $1^{2}B_{2}$ and $3^{2}A^{'}$ states by about 2800 cm$^{-1}$ and 1200 cm$^{-1}$ respectively. Therefore, these molecular states may be considered as the vibrationally excited states of the $1^{2}B_{2}$ and $3^{2}A^{'}$ states. Assuming that the potential energy surface near the minimum can be approximated by harmonic oscillators, the vibrational frequencies of the three normal modes of the $1^{2}B_{2}$ and $3^{2}A^{'}$ states are calculated to be $(v_1,v_2,v_3)=(120.0,104.4,68.8) $ and $(92.0, 78.5, 47.3)$ cm$^{-1}$, respectively.  The energy of a vibrationally excited state may be given by $E=n_1 v_1+n_2 v_2+n_3 v_3$, where $n_1,n_2$, and $n_3$ are the quantum numbers of the three normal modes. The cumulative number of vibrational levels against their energy can be counted and the result is plotted in Fig. 4. Below the spectral region studied in our experiment, there are about 3600 and 600 vibrational levels for the $1^{2}B_{2}$ and $3^{2}A^{'}$ states respectively. The number of states will be larger if the anharmonicity of the potential is considered. For such high-lying states, the assignment of the vibrational quantum numbers for individual molecular states is difficult. We may still compare the density of states which is given by the gradient of the curve. The density of states near the spectral region studied in our experiment is about 4 and 1 per cm$^{-1}$ for the $1^{2}B_{2}$ and $3^{2}A^{'}$ states, respectively. The estimated density of states is higher than the experimental observations, which indicates that some molecular states are not observed, probably because they are very narrow or the Franck-Condon overlap is low.

\begin{figure}[ptb]
\centering
\includegraphics[width=8cm]{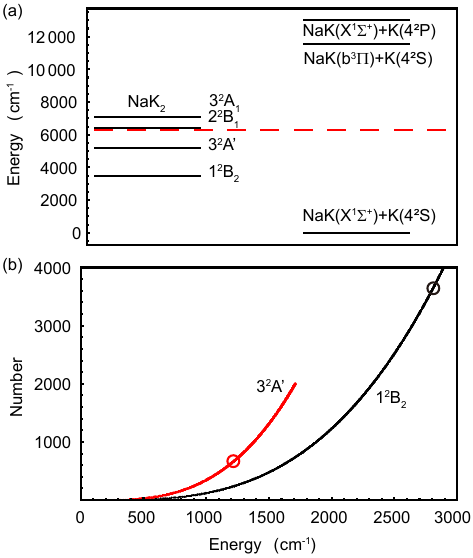}
\caption{The energy of the equilibrium states of the doublet states of $^{23}$Na$^{40}$K$_{2}$ molecules. (a), The four lowest equilibrium states above the NaK(X$^{1}\Sigma^{+})+$ K(4$^{2}$S) dissociation limit obtained by \emph{ab initio} calculations. The energy of the atom-diatomic molecule dissociation limit is also shown. The red dashed line represents the spectral region studied in the current experiment. (b), The cumulative number of vibrational states as a function of their energy  for the $1^{2}B_2$ and $3^{2}A'$ states. The density of states is given by the gradient. The spectral region in the current experiment is marked by circles.}%
\label{fig4}%
\end{figure}

Although the exact vibrational quantum numbers cannot be assigned due to the high density of states, the dependence on the polarization of the photoassociation laser may allow us to assign the rotational quantum numbers.  For triatomic molecules formed by photoassociation of the atoms and diatomic molecules in the maximally polarized state, the nuclear spins may be neglected. In this case, the total angular momentum $J=N+S$ of triatomic molecules may be approximately a good quantum number, where $N$ is the rotational quantum number and $S$ is the electron spin \cite{Wang2021}. The initial scattering state has a total angular momentum $J=n+l+S=1/2$ and the projection along the magnetic field is $M_J=-1/2$, where $n=0$ represents the rotational quantum number of $^{23}$Na$^{40}$K molecules, $l=0$ is the angular momentum of relative motion, and $S=1/2$ is the electron spin of $^{40}$K atoms. According to the selection rule of electronic dipole transition $\Delta J=0,\pm1$, the total angular momentum of the excited molecules may be described by $J'=1/2$ or $3/2$. If $J'=1/2$, because the initial state has $M_J=-1/2$, only the $\sigma^{+}$ transition is allowed, while the $\sigma^{-}$ transition is forbidden. If $J'=3/2$, both transitions are allowed. However, the transition strength of the $\sigma^{-}$ transition is 3 times larger than that of the $\sigma^{+}$ transition. Therefore, for $J'=3/2$, the photoassociation is dominated by the $\sigma^{-}$ transition. According to these arguments, the molecules will be dominantly formed by either $\sigma^{+}$ or $\sigma^{-}$ light. This argument is consistent with the experimental results. Therefore, we assign the rotational quantum number for the nine molecular states. The results are listed in Tab. 1.

In conclusion, we have observed photoassociation resonances in ultracold collisions between between $^{23}$Na$^{40}$K molecules and $^{40}$K atoms. The observation of photoassociation of atoms and diatomic molecules into well-defined quantum states of triatomic molecules paves the way toward the preparation of ultracold deeply-bound triatomic molecules in their ground electronic state. As proposed in Refs. \cite{Olivier2015,Schnabe2021,Shammout2023,Elkamshishy2023}, the ground-state triatomic molecules can be simply produced through spontaneous decay from the excited triatomic molecules. However, the efficiency will be limited by the Franck-Condon overlap between the excited molecular states and the ground states. An efficient and coherent method requires two-photon Raman transfer. The well-resolved excited triatomic molecular state can be used as an intermediate state to search for the deeply bound molecular state in the ground electronic state using two-photon photoassociation.

The photoassociation of atoms and molecules provides a new high-resolution spectroscopy technique for triatomic molecules. Similar to photoassociation spectroscopy of atoms \cite{Stwalley1999,Jones2006}, this technique probes the highly vibrationally excited molecular states, which are difficult to probe using conventional methods that start from the equilibrium configuration due to the low Franck-Condon overlap or symmetry constraints. These highly vibrationally excited molecular states are difficult to understand, owing to the anharmonicity of the interaction potential, large amplitude motions, and the coupling between different modes. Quantitatively understanding these states requires solving a three-body Schr\"{o}dinger equation using a global three-dimensional potential energy surface. Therefore, these high vibrational molecular states provide a valuable benchmark for precisely calculating the potential energy surface and the three-body bound states.

The observation of photoassociation resonances in ultracold atom-diatomic molecule collisions is also important to harness the power of atom-diatomic molecule Feshbach resonances \cite{Yang2019,Son2022}. The photoexcitation of the triatomic collision complex by the trapping light may contribute to a strong loss mechanism in the vicinity of atom-diatomic molecule Feshbach resonance \cite{Christianen2019,Gregory2020,Liu2020,Nichols2022}. The enhanced loss limits the use of Feshbach resonance as a knob to control the atom-diatomic molecule interactions. Our work demonstrates that by tuning the frequency of the trapping light to be far away from a photoassociation resonance, the loss can be suppressed and the lifetime of the mixture can be extended. We may expect the extension of the lifetime can be further improved by using a longer-wavelength trapping laser. The suppression of losses will make the atom-diatomic molecule Feshbach resonance a powerful tool to study strongly interacting quantum gases.

\section*{Methods}

\noindent
\textbf{Experimental sequence}

\noindent
Our experiment starts with the preparation of a quantum degenerate mixture of $^{23}$Na atoms and $^{40}$K atoms with a large number imbalance at a temperature of approximately 100 nK in a large-volume three-beam optical dipole trap at a wavelength of 1064 nm \cite{Cao2023}. At the end of optical evaporative cooling, we load the atoms into a long-wavelength crossed-beam optical dipole trap by slowly ramping the intensity of the trapping light. The wavelengths of the trapping lasers are 1558 nm and 1583 nm, respectively.  The horizontal trapping beam propagates along the direction of the magnetic field. The horizontal trapping beam has a power of about 0.9 W and the beam waist is about 70 $\mu$m. The vertical trapping beam propagates perpendicular to the direction of the magnetic field. The vertical trapping beam has a power of about 0.6 W and the beam waist is about 75 $\mu$m. After loading the atoms into the long-wavelength optical dipole trap, we use the atomic Feshbach resonance between the $|f,m_f\rangle_{\rm{Na}}=|1,1\rangle$ and $|f,m_f\rangle_{\rm{K}}=|9/2,-9/2\rangle$ states at about 78.3 G to create $^{23}$Na$^{40}$K Feshbach molecules by ramping the magnetic field from 81 G to 77.6 G, where $f$ and $m_f$ are the hyperfine quantum number and its projection along the magnetic field, respectively.  The $^{23}$Na$^{40}$K Feshbach molecules are then transferred to the rovibrational ground state of the singlet electronic ground state $X^{1}\Sigma^{+}$ by stimulated Raman adiabatic passage. We prepare $^{23}$Na$^{40}$K molecules in the hyperfine level $|v,n,m_{\rm{Na}},m_{\rm{K}}\rangle=|0,0,-3/2,-4\rangle$ of the rovibrational ground state, where $v$ and $n$ represent vibrational and rotational quantum numbers, and $m_{\rm{Na}}$ and $m_{\rm{K}}$ represent the projections of the nuclear spins. After removing the $^{23}$Na atoms by a resonant laser pulse, we obtain an ultracold mixture of $^{23}$Na$^{40}$K molecules and $^{40}$K atoms in a long-wavelength crossed-beam optical dipole trap.

The Feshbach resonance between $^{23}$Na$^{40}$K molecules and $^{40}$K atoms at about 48.2 G \cite{Wang2021,Su2022,Yang2022b} is used to enhance the free-bound Franck-Condon overlap. After preparing the ultracold atom-diatomic molecule mixture at 77.6 G, we quickly ramp the magnetic field to 49 G. After a controllable hold time, we remove the $^{40}$K atoms by a resonant light pulse. We then ramp the magnetic field back to 77.6 G and transfer the $^{23}$Na$^{40}$K molecules to the Feshbach state for detection. The horizontal trapping laser is used as the photoassociation laser. We monitor the decay of the $^{23}$Na$^{40}$K molecules as a function of the photoassociation laser. The two trapping lasers are both fiber lasers. They have a short-term linewidth of several kHz and a long-term frequency drift of about 20 MHz. The laser frequency can be tuned by up to about 100 GHz via changing the temperature. The frequency of the laser is determined by a wavelength meter with a resolution of 8 MHz.

\bigskip

\noindent
\textbf{Ab initio calculations}

We conduct electronic structure calculations for the doublet $^{23}$Na$^{40}$K$_2$ using the MOLPRO package~\cite{Werner2012}. The calculations are performed on the CASSCF+MRCI level of theory. To model the alkali-metal atoms, we use a single electron model and consider the interaction between the core and valence electrons using an effective core potential (ECP) along with a core polarization potential (CPP). We utilize the ECPxSDF family of core potentials developed by the Stuttgart group~\cite{Fuentealba1982, Fuentealba1983}. These core potentials precisely represent the core-valence interaction, enabling reliable electronic structure calculations for alkali-metal systems. In our calculations, we use uncontracted $sp$ basis sets that are specifically designed for the ECPxSDF core potentials~\cite{Fuentealba1982, Fuentealba1983}. These basis sets are augmented by additional $s$, $p$, $d$, and $f$ functions, each with their respective coefficients~\cite{Ziuchowski2010}. The coefficients for sodium (Na) are: $s: 0.009202$, $p: 0.005306$, $d: 0.3, 0.1$, and $f: 0.1$. The coefficients for potassium (K) are: $s: 0.009433$, $p: 0.004358$, $d: 0.27, 0.09$, and $f: 0.09$. We utilize core polarized potentials based on those of Muller and Meyer~\cite{Mueller1984}. We use the cutoff parameter suggested in Ref.\cite{Ziuchowski2010}: $0.82$ for Na and $0.36$ for K. Combining these basis sets and ECP+CPPs, we compute the energy of the equilibrium states of the doublet electronic excited states of the $\text{NaK}_{2}$ molecule. Specifically, we perform geometric optimization in $C_{2v}$ symmetry to obtain the geometries and energies of the $3^{2}A_1$, $2^{2}B_1$, $1^{2}B_2$ and $2^2A_1$ states. Subsequently, we analyze the vibrational frequencies of these states. The $2^2A_1$ state exhibits imaginary vibrational frequencies, indicating an unstable configuration. Therefore, we further optimize the geometry of this state in $C_s$ symmetry, identifying it as the $3^2A'$ state. The energies of the $3^{2}A_1$, $2^{2}B_1$, $1^{2}B_2$, and $3^2A'$ states are shown in Fig. 4 of the main text.  The calculated vibrational frequencies of the three normal modes of the $1^{2}B_{2}$ and $3^{2}A^{'}$ states are $(v_1,v_2,v_3)=(120.0,104.4,68.8) $ and $(92.0, 78.5, 47.3)$ cm$^{-1}$, respectively.

\renewcommand\subsection[1]{
\vspace{\baselineskip}
\textbf{#1}
\vspace{0.5\baselineskip}}

\subsection{Acknowledgement}

We thank Feng Zhang for helpful discussions. This work was supported by the National Natural Science Foundation of China (under Grant No. 12241409 and 12274393), the National Key R\&D Program of China (under Grant No. 2018YFA0306502), the Chinese Academy of Sciences, the Anhui Initiative in Quantum Information Technologies, the Shanghai Municipal Science and Technology Major Project (Grant No.2019SHZDZX01), the Innovation Program for Quantum Science and Technology (Grant No. 2021ZD0302101).

\subsection{Competing financial interests}

The authors declare no competing financial interests.

\end{document}